\documentclass[aps,prl,reprint,twocolumns,superscriptaddress]{revtex4-2}
\usepackage[utf8]{inputenc}
\usepackage{amsmath}
\usepackage{amssymb}
\usepackage{txfonts}
\usepackage[hidelinks]{hyperref}
\usepackage{tikz}
\usepackage{hyperref}
\usepackage{color}
\hypersetup{colorlinks=true}
\usepackage{graphicx}

\usepackage{bm}

\usepackage{chemformula}
\usepackage[capitalise]{cleveref}
\usepackage{siunitx}

\newcommand{\deltaq}{\eta_{{\mathrm{min}}}(\theta)/\eta_{{\mathrm{maj}}}(\theta)}

\DeclareMathOperator{\Imm}{Im}
\DeclareMathOperator{\Ree}{Re}

\newcommand{\xx}{\hat{\bm{x}}}

\newcommand{\mdos}{\tilde{m}}
\newcommand{\kF}{k_{F}}
\newcommand{\vs}{v_d}

\newcommand{\criticalangle}{\SI{39.9}{\degree}}
\newcommand{\Fermilevel}{\SI{0.5}{eV}}
\newcommand{\mstar}{1.25}
\newcommand{\mzero}{0.4}

\newcommand{\muB}{0.025}
\newcommand{\muBShift}{\SI{1.5}{\mu eV}}
\newcommand{\demonenergy}{\SI{2.8}{meV}}
\newcommand{\muBShiftrelative}{0.1}
\newcommand{\muBmax}{0.1}

\newcommand{\subfigref}[2]{Fig.~\hyperref[#1]{\ref*{#1}#2}}

\newenvironment{DIFnomarkup}{}{}

\begin{document}
	\title{Spin demons in $d$-wave altermagnets}
	\date{\today}
	
	\author{Pieter M. Gunnink}
	\email{pgunnink@uni-mainz.de}
	\author{Jairo Sinova}
	\author{Alexander Mook}
	\address{Johannes Gutenberg University Mainz, Staudingerweg 7, Mainz 55128, Germany}
	\begin{abstract}
		Demons are a type of plasmons, which consist of out-of-phase oscillations of electrons in different bands. Here, we show that $d$-wave altermagnets, a recently discovered class of collinear magnetism, naturally realize a spin demon, which consists of out-of-phase movement of the two spin species.
		The spin demon lives outside of the particle-hole continuum of one of the spin species, and is therefore significantly underdamped, reaching quality factors of $>10$. We show that the spin demon carries a magnetic moment, which inherits the $d$-wave symmetry. Finally, we consider both three and two dimensional $d$-wave altermagnets, and show that spin demons exists in both.
	\end{abstract}
	
	\maketitle

	\paragraph{Introduction.}
	Altermagnets are a recently discovered class of collinear magnets, characterized by a sublattice transposing symmetry involving rotation or mirror operations \cite{smejkalConventionalFerromagnetismAntiferromagnetism2022,smejkalEmergingResearchLandscape2022}. Their anisotropically spin-split Fermi surfaces exhibit a $d$-wave (or higher even-parity) order. These spin-split bands can give rise to unusual transport properties \cite{smejkalEmergingResearchLandscape2022,zarzuelaTransportTheorySpintransfer2024,liaoSeparationInverseAltermagnetic2024}, piezomagnetism \cite{aoyamaPiezomagneticPropertiesAltermagnetic2024,yershovFluctuationinducedPiezomagnetismLocal2024}, the generation of spin-splitter torque in
	MRAM geometries \cite{karubeObservationSpinSplitterTorque2022} and chiral split magnon bands \cite{nakaSpinCurrentGeneration2019,liuChiralSplitMagnon2024, smejkalChiralMagnonsAltermagnetic2023}.
	
	The existence of spin-split Fermi surfaces also opens up the possibility of an out-of-phase oscillation of the two spin densities, realizing a \emph{demon}: an acoustic, electrically neutral type of plasmon, first proposed by \textcite{pinesElectronInteractionSolids1956} in 1956 and only observed by \textcite{husainPinesDemonObserved2023} in 2023. Demons are gapless, in contrast to the conventional in-phase charge plasmon in three dimensions, and have  been predicted for numerous materials \cite{pinesElectronInteractionSolids1956,ruvaldsAreThereAcoustic1981,dassarmaCollectiveModesSpatially1981,sadhukhanNovelUndampedGapless2020,afanasievAcousticPlasmonsTypeI2021}. Importantly, demons are not infinitely long-lived quasiparticles, because they have finite overlap with the particle-hole continuum. This overlap can however be reduced with sufficient separation of the Fermi surfaces, suppressing the damping and thus forming well-defined quasiparticles with high quality factors \cite{agarwalLonglivedSpinPlasmons2014,husainPinesDemonObserved2023}. Besides being relevant as a long-lived quasiparticle, demons have also been proposed to affect phase-transitions \cite{varmaMixedvalenceCompounds1976a}, couple to phonons \cite{sinhaPossibilityAcousticPlasmons1983} and induce superconductivity \cite{ihmDemonsSuperconductivity1981,ruvaldsAreThereAcoustic1981,afanasievAcousticPlasmonsTypeI2021}.

	In this work, we show that in a $d$-wave altermagnetic metal, the spin-split bands naturally host a out-of-phase oscillation of the spin densities---thus realizing a spin-polarized demon, which we dub a \emph{spin demon}. 
	The spin demon does not live completely outside of the particle-hole continuum, but only outside of the particle-hole continuum of one of the spin species. The spin demon is therefore not completely undamped, but can still reach quality factors of $>10$ for realistic parameters and is therefore well defined and long lived. We demonstrate the existence of the spin demon in both a three-dimensional (3D) and two-dimensional (2D) $d$-wave altermagnetic metal. We also establish that the spin demon inherits the $d$-wave symmetry of the altermagnetic order parameter, by demonstrating that it has a finite magnetic moment which changes sign as the demon's propagation  direction is rotated through the altermagnetic spin-split plane. 
	
	The primary signature of the spin demon is a strongly peaked response in the imaginary part of the spin-spin response function, $\Imm[\chi_{S_zS_z}(\bm q,\omega)]$, as shown in the altermagnetic spin-split plane in \cref{fig:polar}. It follows the four-fold rotational symmetry of the $d$-wave altermagnet, vanishing along the high-symmetry axis, where the electron bands are degenerate. In the four different quadrants of the altermagnetic spin-split plane, the majority spin species in this out-of-phase oscillation changes, following the altermagnetic $d$-wave symmetry. 
	
	\begin{DIFnomarkup}
	\begin{figure}
		\includegraphics[width=\columnwidth]{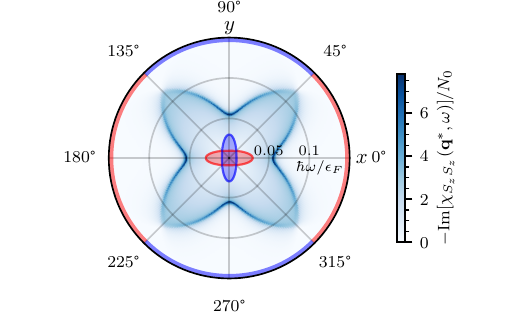}
		\caption{The imaginary part of the spin–spin response function, $\Imm[\chi_{S_zS_z}(\bm q^*,\omega)]$, where $\bm q^*=q\ (\cos\theta,\sin\theta,0)$, rotated in the altermagnetic spin-split plane with a fixed $q=0.05\kF$. The angular direction encodes $\theta$ and the radial axis the frequency $\omega$. The colors on the ring indicate the projected spin species, with red (blue) spin up (down). The spin demon is the sharp resonance that follows the four-fold rotational symmetry of the $d$-wave altermagnet. The anisotropically spin-split Fermi surfaces are schematically shown at the origin (not to scale). \label{fig:polar}	}
	\end{figure}

	\begin{figure}
		\includegraphics{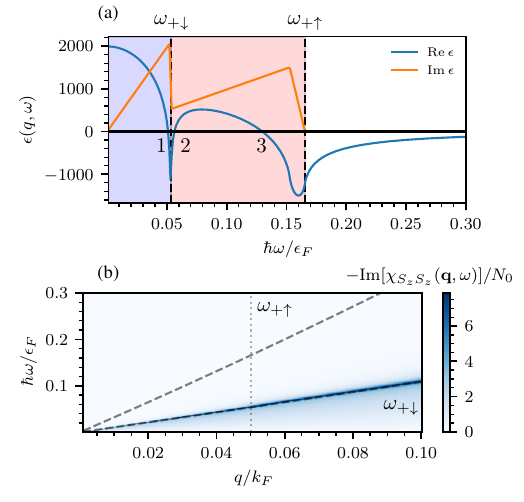}
		\caption{(a) The real and imaginary part of the dielectric function, for $\bm q^* = q \hat{\bm x}$, with $q=0.05\kF$. The zeros of the real part correspond to resonances, the imaginary part determines their damping. The spin demon is the second zero. The blue and red shading indicate where the spin-down and spin-up particle-hole continua is non-zero. (b)  The imaginary part of the spin–spin response function, $\Imm[\chi_{S_zS_z}(\bm q,\omega)]$, for $\bm q=q\xx$, showing the existence of a spin demon with a high quality factor. The vertical dotted line corresponds to the $q$ used in (a). In both (a,b), the dashed lines indicate the spin-resolved particle-hole continua edges, $\omega_{+\sigma}$.\label{fig:alongx} }
	\end{figure}
	\end{DIFnomarkup}


	\paragraph{Method.}
	We describe the spin demon within the random phase approximation (RPA), where the spin-resolved response functions $\chi_{\sigma\sigma'}$ follow \cite{giulianiQuantumTheoryElectron2005}
	\begin{equation}
		\begin{pmatrix}
			\chi_{\uparrow\uparrow} & \chi_{\uparrow\downarrow} \\ 
			\chi_{\downarrow\uparrow} & \chi_{\downarrow\downarrow}
		\end{pmatrix}^{-1} = \begin{pmatrix}
			\chi_\uparrow^{(0)} & 0 \\
			0 & \chi_\downarrow^{(0)}
		\end{pmatrix}^{-1}
		- v_q \begin{pmatrix}
			1 & 1 \\ 1 & 1 
		\end{pmatrix} \label{eq:chi-rpa}.
	\end{equation}
	Here, $\chi_\sigma^{(0)}$ is the non-interacting density-density response function for spin $\sigma$ and $v_q=e^2/\epsilon_0 q^2$ is the Fourier transform of the Coulomb interaction, with $q=|\bm q|$. The (planar) $d$-wave altermagnet is oriented such that the spin-splitting is maximal along the $x,y$-axis, giving the low-energy dispersion \cite{smejkalEmergingResearchLandscape2022}
	\begin{equation}
		\epsilon_{\bm k}^\sigma = \frac{\hbar^2 k^2}{2m_0} + \sigma\frac{\hbar^2 \left( k_x^2-k_y^2\right)}{2m_*} ,
	\end{equation}
	where we take $m_0=\mzero m_e$, $m_*=\mstar m_0$ and a Fermi level of $\epsilon_F=\Fermilevel$ ($m_e$ is the electron mass.)
	The non-interacting density-density response function can  be found analytically from the Lindhard function \cite{ahnAnisotropicFermionicQuasiparticles2021}; we show details in Secs.~I and II in the Supplemental Material (SM) \footnote{See \href{https://arxiv.org/src/2504.11062v2/anc/SM.pdf}{Supplemental Material} under \textit{Ancillary files} for a detailed calculation of the spin resolved response functions, details on the numerics, analytical solutions for the spin demon dispersion, quality factor and magnetic moment, and the spin demon in two dimensions and in $g$-wave altermagnets.}. Solving \cref{eq:chi-rpa} for $\chi_{\sigma\sigma'}(\bm q,\omega)$, we find the three response functions $\chi_{nn(\bm q,\omega)},\ \chi_{nS_z}(\bm q,\omega),\ \chi_{S_zS_z}(\bm q,\omega)$ \cite{giulianiQuantumTheoryElectron2005}. We focus on $\chi_{S_zS_z}(\bm q,\omega)$, which shows the strongest signature of the spin demon:
	\begin{equation}
		\chi_{S_zS_z}(\bm q,\omega)=\frac{\chi_\uparrow^{(0)}+\chi_\downarrow^{(0)}-4v_q\chi_\uparrow^{(0)}\chi_\downarrow^{(0)}}{\epsilon(\bm q,\omega)}
	\end{equation}
	where 
	\begin{equation}
		\epsilon(\bm q,\omega) \equiv 1 - v_q \left(\chi_\uparrow^{(0)}+\chi_\downarrow^{(0)}\right)
	\end{equation}
	is the complex longitudinal dielectric function. We discuss $\chi_{nn}(\bm q,\omega)$ and $\chi_{nS_z}(\bm q,\omega)$ in Sec.~III in the SM \cite{Note1}. 
	
	Collective modes emerge as the poles of the response function, determined by the zeros of the longitudinal dielectric function, 
	\begin{equation}
		\epsilon(\bm q,\omega) = 0, \label{eq:poles}
	\end{equation}
	resulting in a peak in the imaginary part of the response function \cite{giulianiQuantumTheoryElectron2005}. 
	

	We first analyze the dielectric function in more detail, by showing $\epsilon(\bm q,\omega)$ for a fixed $\bm q\parallel \xx$ in \subfigref{fig:alongx}{(a)}. We indicate the spin-polarized particle-hole continua, which edges are given by $\omega_{+\sigma}=\eta_\sigma(\theta) v_F q+ O(q^2/\kF^2)$, where $\eta_\sigma(\theta)$ and $v_F$ are defined in \cref{eq:sigma}.
	
	We observe the existence of three zeros of the dielectric function. The first and third zero correspond to the spin-down and spin-up acoustic plasmon respectively \cite{kamenevFieldTheoryNonEquilibrium2023,kaltenbornPlasmonDispersionsSimple2013}, which are overdamped because they live in their respective particle-hole continuum. The second zero however arises because of the interplay of the spin-up and spin-down particles, and we claim here that this corresponds to a spin demon. We show the evolution of the three zeros in the altermagnetic spin-split plane in Sec.~IV of the SM \cite{Note1}.
	Importantly, the spin demon sits outside of the spin-down continuum, and therefore the imaginary part of the dielectric function is reduced. This implies that the spin demon is potentially underdamped.
	
	We show the underdamped character of the spin demon in more detail with the imaginary part of the spin-spin response function, $\Imm[\chi_{S_zS_z}(\bm q,\omega)]$ in \subfigref{fig:alongx}{(b)}. As noted earlier, collective modes result in peaks in the imaginary part of a response function, with the width of the peak inversely proportional to its lifetime. We observe in \subfigref{fig:alongx}{(b)} a sharply peaked resonance close to the edge of the spin-down continuum---which we identify as the spin demon from our previous analysis of the zeros of $\epsilon(\bm q,\omega)$. We stress that the spin demon is not completely undamped, due to a finite overlap with the spin-up continuum. 
	
	We show in Sec.~V in the SM \cite{Note1} the imaginary part of the spin-spin response function $\Imm[\chi_{S_zS_z}(\bm q,\omega)]$ for a larger wave vector range, from which we conclude that for this set of parameters, the spin demon remains well defined for $q/\kF\lesssim 0.5$.

	Upon rotation through the altermagnetic spin-split plane, we obtain \cref{fig:polar}, demonstrating that the spin demon is most sharply defined along $x$ and $y$, and vanishes along the nodal lines, where the Fermi surfaces are spin degenerate. 
	The spin demon remains well defined for small tilt angles off the altermagnetic spin-split plane, as shown in Sec.~VI in the SM \cite{Note1}.

	\paragraph{Analysis. }
	In what follows, we constrain $\bm q$ to lie in the altermagnetic spin-split plane, parametrizing $\bm q = q\  (\cos\theta,\sin\theta,0)$. Our analysis is simplified by defining the projected spin splitting of the particle-hole continuum for spin species $\sigma$:
	\begin{equation}
		\eta_{\sigma}(\theta) \equiv \sqrt{\mdos m_0\left( 1+\sigma \frac{m_0}{m_*}\cos2\theta\right)} \label{eq:sigma},
	\end{equation}
	where $\mdos\equiv m_0({m_*^2}/({m_*^2-m_0^2}))^{1/3}$.
	We have defined $\eta_{\sigma}(\theta)$ such that $\chi^{(0)}_\sigma$ can be obtained from the well-known Lindhard function for spherical Fermi surfaces \cite{lindhardPropertiesGasCharged1954,giulianiQuantumTheoryElectron2005} by rescaling $q\rightarrow\eta_\sigma(\theta) q$ and $m\rightarrow \mdos$ \cite{ahnAnisotropicFermionicQuasiparticles2021}. For convenience, we define a spin-independent Fermi wave vector $\kF\equiv\sqrt{2\mdos E_F}/\hbar$ and velocity $v_F\equiv \hbar k_F/\mdos$.
	
	The analysis is simplified by noting that, depending on the angle $\theta$, one of the two spin species can be treated as the (projected) majority spin species, defined such that $\eta_{{\mathrm{maj}}}(\theta)>\eta_{{\mathrm{min}}}(\theta)$. For example, along $x$, spin down is the minority spin species (cf. \cref{fig:alongx}).
	We solve for the zero in the dielectric function corresponding to the spin demon by making the ansatz \cite{santoroAcousticPlasmonsConducting1988}
	\begin{equation}
		\omega_{d}(\bm q)=\vs \eta_{{\mathrm{min}}}(\theta)q
	\end{equation}
	and requiring $\omega_{d}(\bm q)$ to lie in the pseudogap formed by the edges of the spin-resolved particle-hole continua.
	
	We carry out this approach in Sec.~VIII in the SM \cite{Note1}, and find that, up to corrections of order $(q/\kF)^2$, the spin demon velocity $\vs$ is determined by 
	\begin{equation}
		4 -\frac{\vs}{v_F} \log\left[\frac{v_F-\vs}{\vs-v_F}\right] - \frac{\vs\eta_{{\mathrm{min}}}}{v_F\eta_{{\mathrm{maj}}}} \log\left[\frac{v_F\eta_{{\mathrm{maj}}}-\vs\eta_{{\mathrm{min}}}}{\vs\eta_{{\mathrm{min}}}-v_F\eta_{{\mathrm{maj}}}}\right] = 0. \label{eq:zeros-vs}
	\end{equation}
	This has no analytical solutions, and we thus solve it numerically.


	The above analysis also gives the quality factor, defined as $Q \equiv \omega_d / \gamma$, where the damping $\gamma$ can be obtained by performing a Laurent-Taylor expansion around $\omega_d$ to find 
	\begin{equation}
		\gamma = \frac{\Imm[\epsilon(\bm q,\omega)]}{\partial_\omega \Ree[\epsilon(\bm q,\omega)]}\Bigr|_{\omega=\omega_d}.
	\end{equation}

\begin{figure}
		\includegraphics{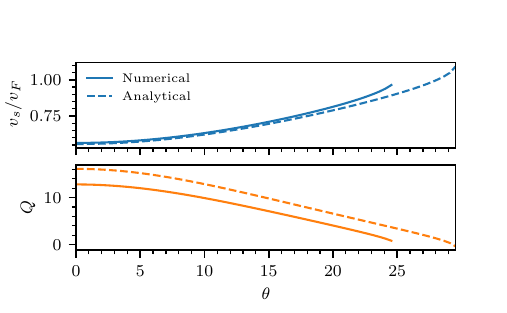}
		\caption{The spin demon velocity (top) and quality factor (bottom), as a function of angle $\theta$. The numerical solutions (solid),  are obtained by numerically finding the zeros and corresponding derivatives from the full dielectric function; the analytical solutions (dashed) follow from numerically solving \cref{eq:zeros-vs}. 
			\label{fig:vs-and-Q} }
	\end{figure}

	In \cref{fig:vs-and-Q}, we show $\vs$ and the corresponding quality factor $\gamma$ as a function of $\theta$. We stress that for quality factors less than unity, the spin demon is no longer a well-defined quasiparticle, which happens for $\theta_c\ge\SI{23}{\degree}$ for this set of parameters. Up to this critical angle, the velocity of the spin demon only changes by a factor of 2, while the quality factor falls off by one order of magnitude. The quality factor is not bounded, and increasing the altermagnetic band anisotropy (proportional to $m_0/m_*$) leads to higher quality factors.
	
	
	\paragraph{Out-of-phase oscillations and magnetic moment. }To gain more insight in the character of the spin demon, we solve the eigenvalue problem defined by \cref{eq:chi-rpa}, 
	\begin{equation}
		\begin{pmatrix}
			\Ree[\chi_{\uparrow}^{(0)}(\omega)]^{-1}-v_q & -v_q \\
			-v_q & \Ree[\chi_{\uparrow}^{(0)}(\omega)]^{-1}-v_q
		\end{pmatrix}\begin{pmatrix}
			\psi_\uparrow \\ \psi_\downarrow
		\end{pmatrix} =0,
	\end{equation}
	which has the solution
	\begin{equation}
		\frac{\psi_{{\mathrm{maj}}}}{\psi_{{\mathrm{min}}}} = -\frac{v_qN_0}{1 + v_qN_0}\approx -1 + O(q^2/\kF^2) \label{eq:out-of-phase},
	\end{equation} 
	where $N_0=\mdos \kF / (2\pi^2\hbar^2)$ is the spin-independent density of states at the Fermi level.
	This result thus clearly shows that in the limit of small $q/\kF$, the spin demon consists of out-of-phase oscillations of two spin-species---in contrast to the conventional plasmon, which consists of in-phase oscillations \cite{agarwalLonglivedSpinPlasmons2014}. \Cref{eq:out-of-phase} also demonstrates that as $q/\kF$ approaches zero, $|{\psi_{{\mathrm{maj}}}}| < |{\psi_{{\mathrm{min}}}}|$. We therefore expect that a spin demon carriers a magnetic moment, since it is composed of predominantly one spin species. 
	
	To show this in more detail,  we consider an external magnetic field $B$ aligned with the N\'{e}el vector direction and with a magnitude far below the spin-flop transition. The electrons  gain energy $\sigma g_e\mu_B B$, with $g_e\approx2$ the electron gyromagnetic ratio and $\mu_B$ the Bohr magneton. We furthermore neglect orbital magnetization effects. We now calculate the magnetic moment, which is defined as
	\begin{equation}
		\mu_d \equiv -\hbar\frac{\partial\omega_d}{\partial B}. \label{eq:mud}
	\end{equation}
	In the limit of $\deltaq\rightarrow0$ we have
	\begin{equation}
		\frac{\partial \vs}{\partial B} = \vs' \frac{\partial\Delta}{\partial B} + O(\Delta^2),
	\end{equation}
	where $\Delta\equiv g_e\mu_B B N'_0 /N_0 $, $ \vs'=8e^{-4}v_F$ and $N'_0=\partial N_0(\epsilon)/\partial\epsilon|_{\epsilon=\epsilon_F}$, 
	This allows us to obtain the magnetic moment as
	\begin{equation}
		\mu_d =  g_e\mu_B\hbar \frac{N'_0}{N_0}  \eta_{\mathrm{min}}(\theta)  \vs' q. \label{eq:mup}
	\end{equation}
	Importantly, a finite magnetic moment implies that a spin demon can carry angular momentum. 
	In addition, since the minority spin species switches between spin-up and spin-down as the spin demon is rotated through the plane, the magnetic moment has the opposite sign for $\bm q \parallel \hat{\bm y}$, representing the $d$-wave symmetry of the underlying altermagnetic bandstructure. 
	
	\begin{figure}
		\centering
		\includegraphics[width=0.8\columnwidth]{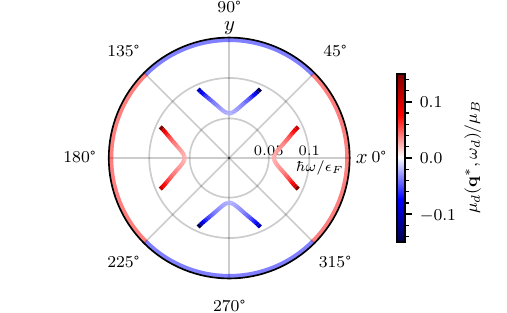}
		\caption{Numerically obtained spin demon resonance frequency $\omega_d$ as a function of $\bm q^*=q\ (\cos\theta, \sin\theta,0)$, with $q=0.05\kF$, rotated in the altermagnetic spin splitting plane with angle $\theta$. The color corresponds to the magnetic moment, showing the $d$-wave character. Obtained by numerically finding the zeros and evaluating \cref{eq:mud} from the full dielectric function.
			Along $x$ and $y$, the magnetic moment is approximately $\pm\muB\mu_B$. The colors on the ring indicate the projected spin species. \label{fig:magnetic-moment} }
	\end{figure}
	
	We show this in \cref{fig:magnetic-moment}, where we have numerically calculated the magnetic moment of the spin demon as a function of $\theta$. For the angles where the spin-up species is the majority species, we obtain a positive magnetic moment, whereas for the angles where the spin-down species is dominant, we have a negative magnetic moment. The magnetic moment thus captures the $d$-wave symmetry of the underlying altermagnetic bandstructure [\cref{eq:mup}]. For $q=0.05\kF$, we obtain that $\mu_p\approx\muB\mu_B$ for $\bm q\parallel \xx$. The magnetic moment grows to $\muBmax\mu_B$ for angles approaching the critical angle---but the quality factor also decreases. 
	
	These results show that an applied magnetic field will shift the spin demon frequencies up or down, depending on the orientation of $\bm q$. The shift in the spin demon frequency is however small, approximately \muBShift at $\bm q=0.05k_F\hat{\bm x}$ with a magnetic field of \SI{1}{T}, whereas the energy of the spin demon is \demonenergy, resulting in a relative shift of $\muBShiftrelative\%$. Larger relative shifts might be realized by strain through piezomagnetism \cite{aoyamaPiezomagneticPropertiesAltermagnetic2024}.

	\paragraph{Two-dimensional.}
	The spin demon also exists in two-dimensional altermagnets. The analysis in 2D is similar to in 3D, and we relegate details to Sec.~IX in the SM \cite{Note1}. We choose the same parameters as in 3D.
	
\begin{figure}
		\includegraphics{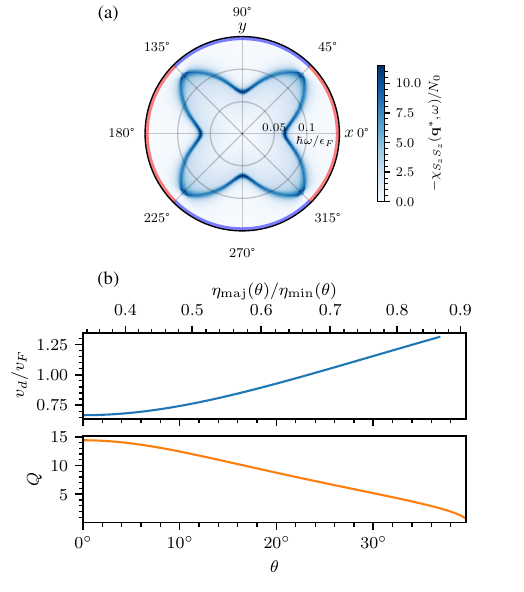}
		\caption{In two dimensions: (a) $\Imm[\chi_{S_zS_z}(\bm q^*,\omega)]$, The angle indicates $\theta$ and the radial axis the frequency $\omega$. (b) The spin demon velocity (top) and quality factor (bottom) as a function of rotation angle  $\theta$. \label{fig:2D} }
	\end{figure}

	We show the resulting the imaginary part of the spin-spin response function in \cref{fig:2D}, highlighting the same four-fold rotational symmetry. 
	In addition, we show the spin demon velocity and quality factor as a function of the projected spin splitting in 2D. Because the particle-hole continuum is sharply defined in two dimensions,  we are able to provide analytical solutions of the spin demon velocity and quality factor as \cite{agarwalLonglivedSpinPlasmons2014}
	\begin{align}
		\vs^{\mathrm{2D}} &= \frac{2}{\sqrt{3}} v_F \eta_{\mathrm{min}}(\theta) q + O(q^2/\kF^2)\\
		Q^{\mathrm{2D}} &= \frac{3\sqrt{4\eta_{\mathrm{min}}^2(\theta)-3\eta_{\mathrm{maj}}^2(\theta)}}{\eta_{\mathrm{min}}(\theta)} + O(q^2/\kF^2),
	\end{align}
	for $2\eta_{\mathrm{min}}(\theta)>\sqrt{3}\eta_{\mathrm{maj}}(\theta)$, whereas the spin demon ceases to exist if this condition is not met.

	We observe that the spin demon in two dimensions is more robust than in 3D, surviving for larger $\theta$ ($\criticalangle$ versus $\approx\SI{23}{\degree}$). Beyond this angle we observe the remnants of the overdamped conventional acoustic plasmon. The quality factors are however comparable in magnitude, especially for angles that align with the altermagnetic axis. Finally, we comment that in 2D, the spin demon also has a magnetic moment (shown in Sec.~IX in the SM \cite{Note1}), which is comparable in magnitude to the 3D case and displays the same altermagnetic symmetry.
	
	\paragraph{Conclusion. }We have shown that both three and two dimensional altermagnetic metals can host out-of-phase oscillations of the two spin densities, realizing a spin demon. 
	The spin demon has a magnetic moment, which changes sign for propagation along different angles through the altermagnetic plane, inheriting the altermagnetic $d$-wave symmetry.
	We have considered only the RPA, since corrections to the RPA have been shown to mainly enhance the damping of comparable acoustic plasmons in a two-dimensional spin-polarized electron gas \cite{kreilExcitationsSpinpolarizedTwodimensional2015}. Furthermore, Fermi liquid descriptions of demons have shown no considerable differences compared to the RPA approach \cite{afanasievAcousticPlasmonsIsotropic2022}. We expect similar conclusions to hold for altermagnetic spin demons.

	In this work, we have considered a $d$-wave altermagnet, where the spin-split Fermi surfaces are elliptical. We have repeated the same analysis for a $g$-wave altermagnet in Sec.~X in the SM \cite{Note1}, where we find that the separation of the spin-polarized particle-hole continua is not sufficient for a spin demon to emerge. We therefore conclude that the spin demon is a specific feature of $d$-wave altermagnets.

	The spin demon could be directly observed by making use of spin-sensitive electron scattering probes, such as spin-polarized electron energy loss spectroscopy (SPEELS) \cite{plihalSpinWaveSignature1999} or cross-polarized Raman scattering \cite{kimPolarizedRamanSpectroscopy2020}. These probes directly measure $\Imm[\chi_{S_zS_z}(\bm q,\omega)]$ (or $\Imm[\chi_{nS_z}(\bm q,\omega)]$, which also contains information of the spin demon; see Sec.~III in the SM \cite{Note1}), and can thus map out \cref{fig:polar} and \subfigref{fig:alongx}{(a)}. SPEELS is typically resolution limited in the \SIrange{10}{100}{meV} scale  \cite{vasilyevDesignPerformanceSpinpolarized2016, zakeriGenerationSpinpolarizedHot2021}. The spin demon exists up to energies of half the Fermi energy $\epsilon_F$, as shown in Sec.~VI in the SM \cite{Note1}, where the Fermi level of typical $d$-wave altermagnets is of the order \SI{500}{meV} \cite{smejkalEmergingResearchLandscape2022}. This would put the spin demon within the resolution range of SPEELS. 
	
	Real samples will most likely consist of multiple magnetic domains with different orientations of the N\'eel vector. We expect that this will not be a difficulty for the detection of the spin demon, since typical domain sizes in altermagnets can be in the micrometer range \cite{aminNanoscaleImagingControl2024a}, placing an upper limit on the spin demon wavelength of micrometers. A probe which is spatially localized on this length scale can therefore directly detect spin demons. In addition, recent transport experiments have measured a finite anomalous Hall effect signal, demonstrating that altermagnetic domains are not equally populated \cite{jeongMetallicityAnomalousHall2025a,leiviskaAnisotropyAnomalousHall2024, reichlovaObservationSpontaneousAnomalous2024a}, and thus even a spatially delocalized probe could detect spin demons.
	
	The data that support the findings of this article are openly available \cite{gunninkPgunninkPlasmonsaltermagnet2025}.

	\begin{acknowledgments}
		We thank Khalil Zakeri for insightful discussions. This work is in part funded by the Deutsche Forschungsgemeinschaft (DFG, German Research Foundation) -- Project No.~504261060 (Emmy Noether Programme) and TRR 173 -- 268565370 (project A03 and B13). P.~G. acknowledges financial support from the Alexander von Humboldt postdoctoral fellowship. 
	\end{acknowledgments}

\end{document}